RESEARCH ARTICLE                                                            OPEN ACCESS

# Simulation Model Of Functional Stability Of Business Processes

Yuri Monakhov*, Olga Fayman**
*,**(Department of Informatics and IT Security, Vladimir State University, Russian Federation)

**ABSTRACT**
Functioning of business processes of high-tech enterprise is in a constant interaction with the environment. Herewith a wide range of such interaction represents a variety of conflicts affecting the achievement of the goals of business processes. All these things lead to the disruption of functioning of business processes. That's why modern enterprises should have mechanisms to provide a new property of business processes – ability to maintain and/or restore functions in various adverse effects. This property is called functional stability of business processes (FSBP). In this article we offer, showcase and test the new approach to assessing the results of business process re-engineering by simulating their functional stability before and after re-engineering.

## I. INTRODUCTION

The modern enterprises are forced to improve their action constantly. It requires the development and implementation of more effective methods of managing.

A possibility opens to improve the performance of an enterprise if there exists a model of it in the form of interconnected purpose-oriented business processes. The basic approaches to design and re-engineering of business processes are fully presented in literature. Some scholars [1] make attempts to introduce metrics and meters of key functioning indicators of business processes. Some authors [2] propose to control modes of the business processes using simulation, though they don't propose the algorithms and the models that allow making a conclusion about the ability of business processes to work in the intended mode of functioning without failure, however, this direction plays an important role in achieving the business goals.

Functioning of business processes of high-tech enterprise is in a constant interaction with the environment. Herewith a wide range of such interaction represents a variety of conflicts affecting the achievement of the goals of business processes. All these things lead to the disruption of functioning of business processes. That's why modern enterprises should have mechanisms to provide a new property of business processes – ability to maintain and/or restore functions in various adverse effects. This property is called functional stability of business processes (FSBP).

We group the parameters that constitute FSBP into structural, organizational and legal groups.

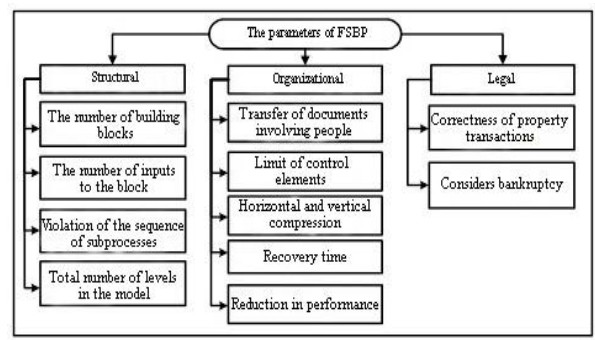

Figure 1 - the parameters of FSBP

As a first step towards the provision of sufficient level of FSBP the current level of functional stability should be estimated.

In this paper we propose the model of FSBP assessment based on the calculation of functional stability indicators. These are represented by a triple:

$$FS = <S, O, L> \qquad (1)$$

S – vector of structural functional stability indicators determining the structural openness of business processes to the external environment.

O – vector of organizational functional stability indicators corresponding to the quantitative estimate of the performance of administrative structures in the conditions of fuzzy human behavior and uncertainty of the environment.

L - vector of legal functional stability indicators determining the stability of business processes in terms of legislation and financial reporting.

Each indicator has its own justification, the range of values, evaluation mechanism and the approach to increasing FSBP level using the increase in this parameter.

FSBP modeling is necessary to assess its characteristics that are reflected in the form of indicators of destabilizing criteria.

It is impossible to assess the presence and the impact of all criteria on FSBP using only one simulation model. That's why it is necessary to create a separate model for each of criteria or for it's group.

In this paper we present the results of FSBP modeling based on the following parameters: "The number of inputs to the block", "The number of building blocks", "Violation of the sequence of subprocesses", "Limit of control elements", "Recovery time".

## II. SIMULATION

Business process is represented as a queuing network. Subprocess is the service channel with the flow of requests (documents, workflow) at the entrance to the block. This flow of requests is modeled as a queue consisting of tasks to be performed by the said subprocess and delays that simulate the processing of the tasks.

For simulation we choose the period of one year. The unit of model time equals one business day. Restrictions on the length of the queue are possible, it depends on the simulated parameters and specifics of the business processes.

Delay time – the approximate time required to process the single workflow instance (i.e. a document).

Each employee processes one request in one time. If several employees working on the same block in the business process, then the number of requests processed per time accordingly increases and it is reflected in the parameter "capacity".

The simulation experiment has a purpose of observing the behavior of the process while varying levels of each functional stability indicator.

To evaluate the functional stability of the specific business process it is necessary to analyze the following indicators after the simulation:

1) Average queue length;

2) The number of dropped requests (requests removed by time-out and requests that were not processed)

3) Utilization of the block

4) Average request processing time (for the whole system)

*FSBP model for parameters "The number of inputs to the block", "The number of building blocks":*

The range of values of the parameter "The number of inputs to the block" is represented in Table 1.

Table 1 - The range of values of the parameter "The number of inputs to the block"

| The parameter value | FSBP level for a given value of the parameter | |
|---|---|---|
| | Quantitative value | Linguistic meaning |
| [1;3] | 0,5 | Not appropriate |
| [3;5] | 1 | Optimal |
| [5;8] | 0,7 | Redundant |
| >8 | 0,2 | Unacceptable |

To increase the FSBP level using the increase in this parameter the redundancy of inputs and outputs should be analyzed. This analysis is performed by reducing the number of inputs and outputs of the business processes by avoiding certain types of inputs and outputs and grouping it into the blocks.

The range of values of the parameter "The number of building blocks" is determined by the SADT methodology [4]. These techniques impose a limit on the number of blocks at each level of decomposition (the rule of 3-6 blocks – power limit of human short-term memory). The lower boundary is chosen because it is not necessary to detail the process by diagram that has one or two processes. The upper limit corresponds to the human possibilities of simultaneous perception and understanding of the structure of a complex system with a lot of internal links. Essence of the evaluation mechanism is in the estimation of the number of blocks in each business process and determining the FSBP level by the expert.

Thus, the quantitative value of the FSBP level at a certain parameter is a re-engineering criterion.

To illustrate the efficiency of the FSBP model (the effect of the FSBP parameters on the parameters of the real business process execution) the simulation of "Development of the product concept" business process of typical high-tech enterprise was performed.

Business process diagram is represented in Figure 2.

Queue length is not limited, because in a real company the information will be accumulated and stored until time when it will be processed or will lose relevance. For this case we add a timeout in the model, after which the information to be processed becomes irrelevant and must make way for more recent information.

The structure of simulation model of the business process "artificially is" for structural parameters "The number of inputs to the block" and "The number of building blocks" run under AnyLogic is represented in Figure 3.

The input data for the model of business process "as is" are represented in Table 2.

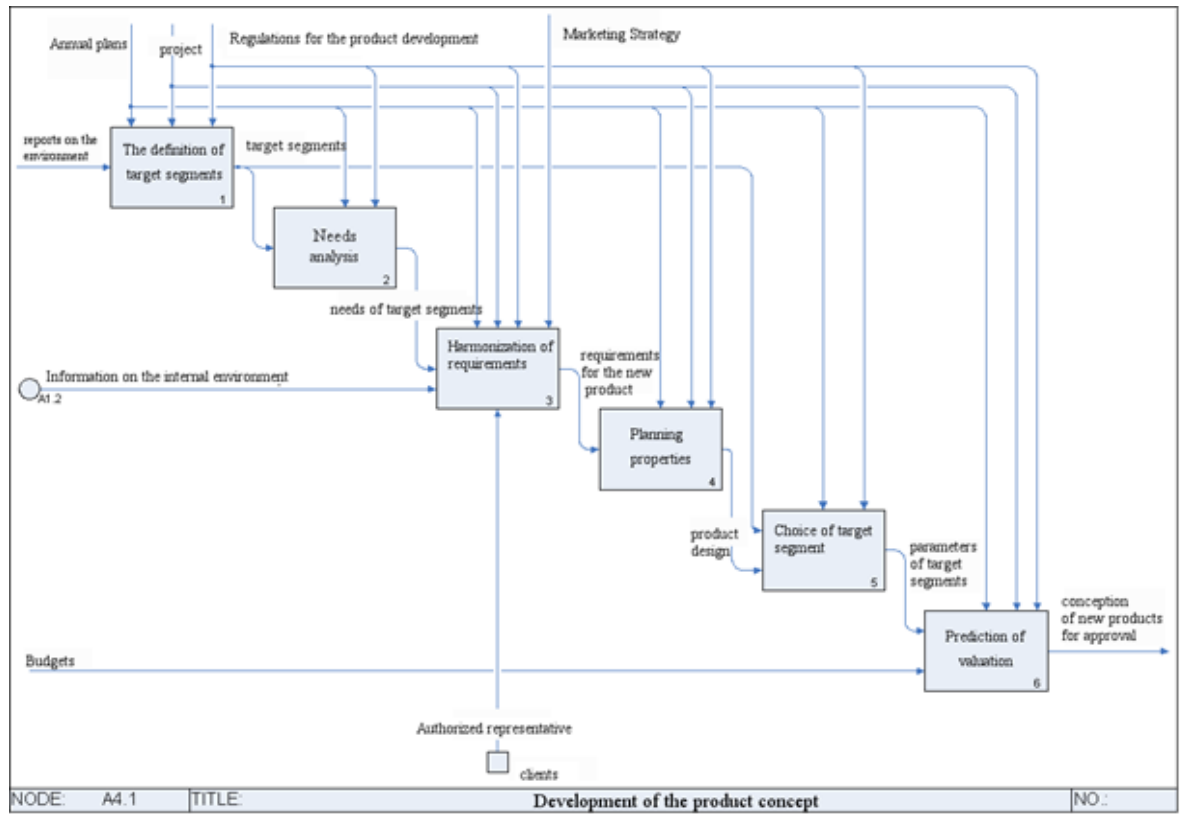

Figure 2 - Business process "Development of the product concept" diagram

Table 2 - The input data for the model of business process "as is".

| The nature of the input requests | Arrival intensity | Restrictions |
|---|---|---|
| Reports on the environment | 1.7 | No restrictions |
| Annual plans | 0.2 | 100 requests for one short period |
| Project | 0.2 | No restrictions |
| Regulations for the product development | 0.2 | 70 requests |
| Marketing strategy | 0.1 | 90 requests |
| Information on the internal environment | 0.7 | No restrictions |
| Authorized representative | 0.7 | 50 requests |
| Budgets | 0.7 | No restrictions |

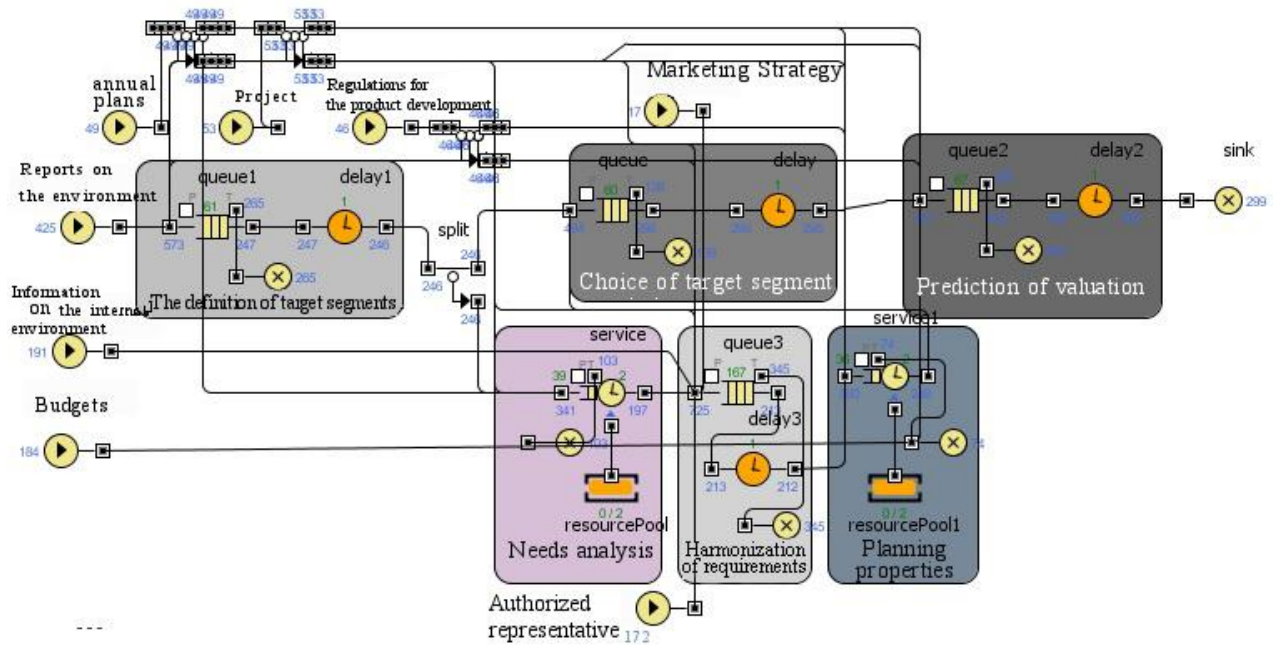

Figure 3 - The structure of simulation model of business process "as is" for structural parameters "The number of inputs to the block" and "The number of building blocks"

During the reengineering of this business process we conclude:

1) Within the framework of the process of six blocks two of them have the increased number of inputs. It is the block number 3 - "Harmonization of requirements" (the number of inputs – 6) and the block number 6 - "Prediction of valuation" (the number of inputs – 5).

2) Blocks number 5 - "Choice of target segment" and number 1 - "The definition of target segments" are mutually redundant, thus indicating that block number 5 is not appropriate to use.

The structure of simulation model of business process "as will be" for structural parameters "The number of inputs to the block" and "The number of building blocks" with the results of re-engineering taken into account is represented in Figure 4.

Table 3 - summarizes the comparative simulation results

| Subprocesses | Efficiency of the channel | | | Number of dropped requests on timeout | | |
|---|---|---|---|---|---|---|
| | Before | After | % | Before | After | % |
| The definition of target segments | 0,995 | 0,997 | Unchanged | 102 | 110 | Increased by 7% |
| Needs analysis | $9,23*10^{-10}$ | $9,23*10^{-10}$ | Unchanged | 22 | 21 | Unchanged |
| Harmonization of requirements | 0,997 | 0,998 | Unchanged | 216 | 231 | Increased by 6% |
| Planning properties | $9,23*10^{-10}$ | $9,26*10^{-10}$ | Unchanged | 74 | 84 | Increased by 12% |
| Choice of target segment | 0,993 | ---------- | | 136 | ------- | |
| Prediction of valuation | 0,997 | 0,994 | Decreased by 1% | 117 | 80 | Decreased by 32% |

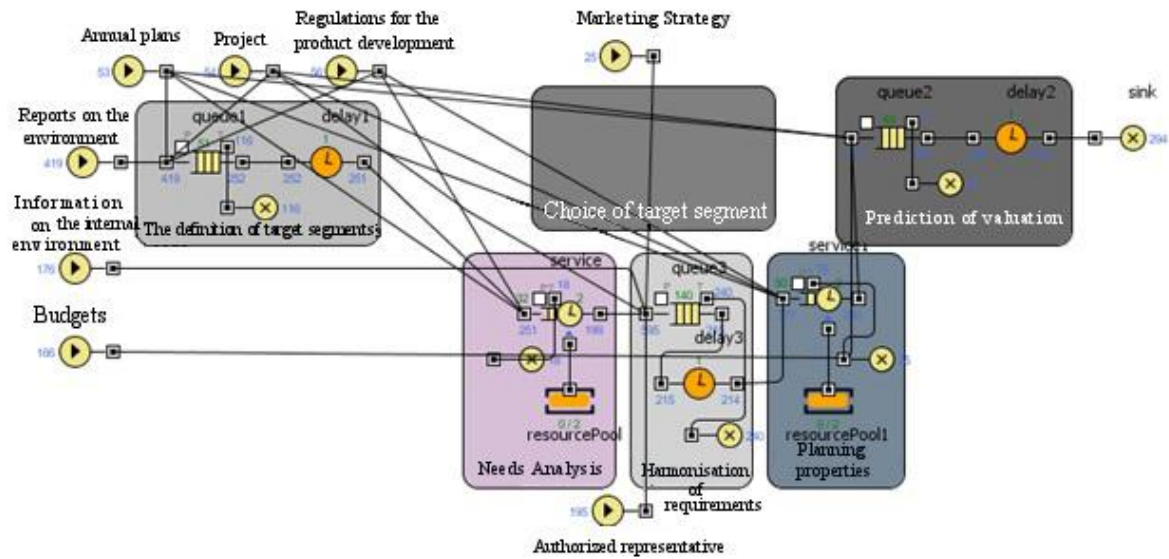

Figure 4 - The structure of simulation model of business process "as will be" for structural parameters "The number of inputs to the block" and "The number of building blocks" with the accounting of the results of re-engineering.

In this research the variation of the average queue length – the metric that defines the actual performance of the business process - was measured (Figure 5).

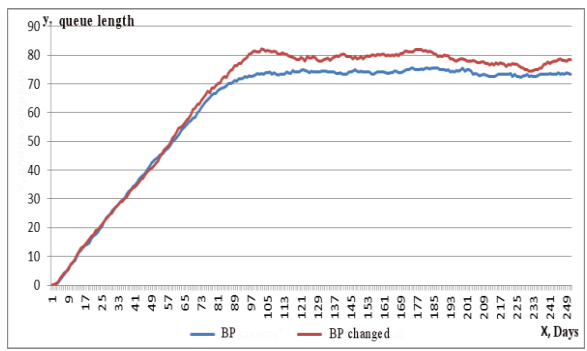

Figure 5 – the variation of the average queue length

During the simulation we increase the FSBP level through the increase of the parameters "The number of inputs to the block" and "The number of building blocks". We observe the following changes of characteristics of the real execution of business process:

1) The number of dropped requests had increased by 7% in the block "The definition of target segments", by 6% in the block "Harmonization of requirements", by 12% in the block "Planning properties" and decreased by 32% in the block "Prediction of valuation";

2) The average request time in the system had decreased by 6% due to the removal of one of the blocks;

3) The queue length had increased by an average of 5 to 10%. This is due to the redistribution of requests that previously were used for the block "Choice of target segment"

4) Utilization of the block remains unchanged.

## III. SIMULATION (CONTINUED)

It is commonly known that the control is the most crucial and important part of any process. Often the number of the control elements is more than necessary. It affects the length of the process, greatly increasing it, therefore reducing the stability of performance.

Some checks in order to reduce the number of the control elements can be carried out by the staff themselves, that is normal at the contemporary level of their skill.

Certainly, the faster checks and control actions will transpire, the cheaper and more stable in terms of accessibility the process will be. Another perspective shows it is important to know the limit in excess of which the speed of the checks will affect the quality of information we receive in the output of business process or its block.

Therefore to ensure the functional stability in terms of character of control elements it is necessary:

1) to define sub processes in which employees can perform the control themselves;

2) to define the number of the external control elements;

3) to define the quality, that is, the speed of the external control elements.

Within the simulation the following restrictions were established:

The length of the queue and the request time in the system were explicitly defined. These parameters are individual and they are set accordingly to the requirements of a specific enterprise for a specific business process.

These parameters have been defined in such a way that the obtained level of functional stability has no negative impact on the business process.

Business process "Production support and quality control" was chosen to showcase the criterion "Limit of control elements" (Figure 6). Figure 7 shows the structure of simulation model for structural parameter "Limit of control elements". The input data of the model are represented in Table 4.

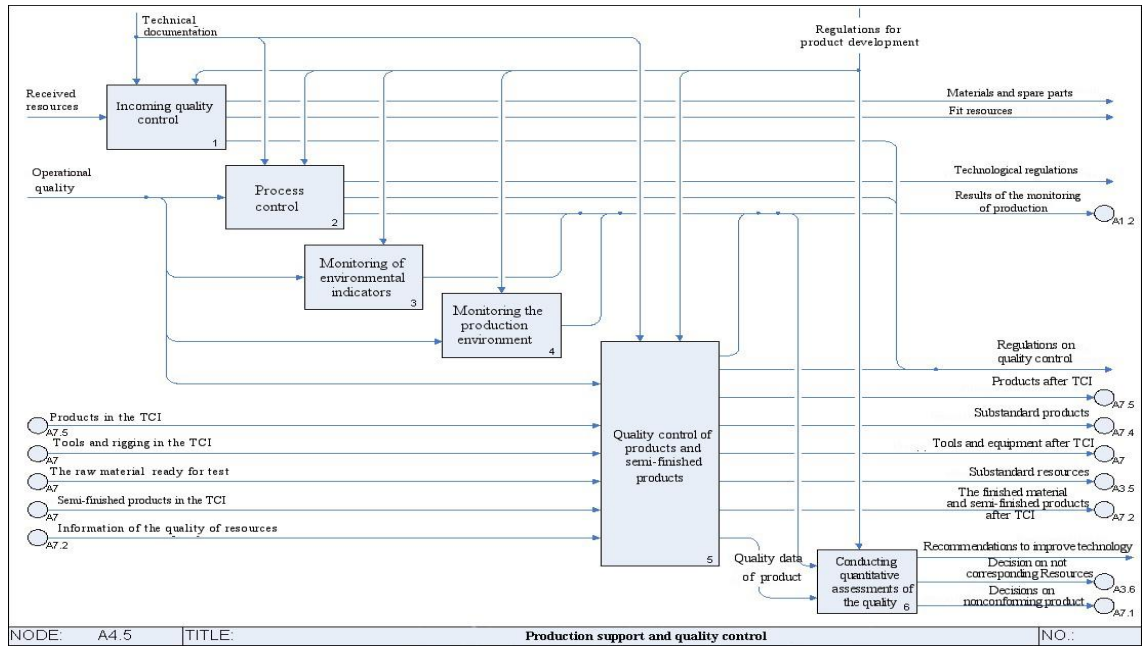

Figure 6 - Business process "Production support and quality control"

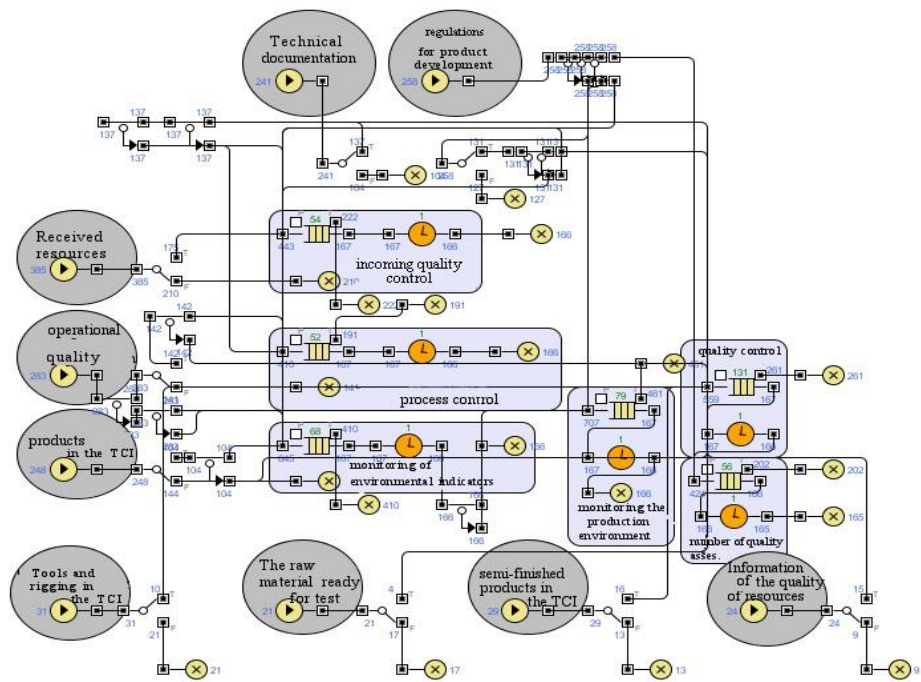

Figure 7 - the structure of simulation model for structural parameter "Limit of control elements"

Table 4 – Input data

| Source | μ | Description |
|---|---|---|
| Received resources | 1 | Resources are received without restrictions with a constant sufficiently high intensity |
| Technical documentation | 0 | It is a large set of various technical documents that are necessary for production. This information is constantly used for production safety. |
| Regulations for product development | 0 | These documents are extensive and detailed, their number is known in advance. They are an important part of any technological process, that's why from time to time it is necessary to access such information. |
| Operational quality | 0 | They act periodically with a specified time interval. |
| Products in the TCI | 1 | The number of requests is constant and strictly regulated. |
| Tools and rigging in the TCI | 0 | The number of requests is less than for the products or raw materials, but it is constant and strictly regulated. |
| The raw material ready for test | 1 | The number of requests is constant and strictly regulated. |
| Semi-finished products in the TCI | 1 | The number of requests is constant and strictly regulated. |
| Information of the quality of resources | 0 | Most of that information is small volumes of statistic data. They are received not so often |

The following values of parameters of control were taken into account during simulation: 0.1, 0.5 and 0.9. It means that in case of 0.1 the regulatory authorities quickly process the requests but with the bad quality. In case of 0.5 the regulatory authorities process the requests with the medium speed and the medium quality level. In case of 0.9 the regulatory authorities process the requests extremely slowly but with the perfect quality.

By analyzing the simulation results, we can see that at 0.1 the load on the control blocks is minimal, there is no queues, there is no dropped requests (displaced by time-out), but all the requests (with good and bad quality) go on for further processing. At 0.5 utilization of the blocks increases significantly, queue and dropped requests appear. At 0.9 utilization of the blocks is close to the limit, queues are too long, and only perfectly correct requests go on for further processing. It takes a lot of time and the most part of potentially correct requests are displaced by time-out. In this case it is important to find such combination of parameters for blocks in which the process will be the most stable. It means that the most part of correct requests will go on for further processing, the invalid requests will be rejected and all of these actions will be performed within the allowable time and resources.

## IV. SIMULATION (CONTINUED)

*FSBP model for parameter "Violation of the sequence of subprocesses".*

To study this parameter the business process "Delivery of goods" was chosen. The initial model is shown in the figures 8 and 9. For simulation the following parameters are used:

Table 5 – The initial parameters of the model

| Source | μ | Description |
|---|---|---|
| Information on product requirements | 1 | Resources are accessed with a constant rate. Critical information quickly becomes obsolete, needs constant updating. |
| Customer request | 1 | Uncontrolled flow of requests is not limited in amount. |
| Prices | 0 | Stable information that is not updated frequently |
| Receivables | 0 | An important information that comes in a small amount and requires periodic updating |
| Competitive information | 1 | A large amount of requests with a small amount of provided information. It requires constant updating. |
| Head of purchase and marketing department | 0 | In theory, it is an important element. In practice this element is used briefly and not often. |
| Job descriptions and specifications | 0 | Stable information, that is updated rarely. Usually it is used in the beginning of the process. |

In this business process you can find some elements that are incorrectly placed in the sequence.

a) The process "assortment planning" should follow the process "forming the request" because the assortment planning should be done on the basis of the existing requests for supply and newly-formed requests in this business process.

b) The process "quality control of delivery" should be performed in parallel with the processes "Gathering information about the suppliers", "purchase of goods", "delivery of goods" and "assortment planning", providing the constant monitoring of quality of the customer service during the whole business process.

We will change the model to correct current deficiencies. (Figure 10)

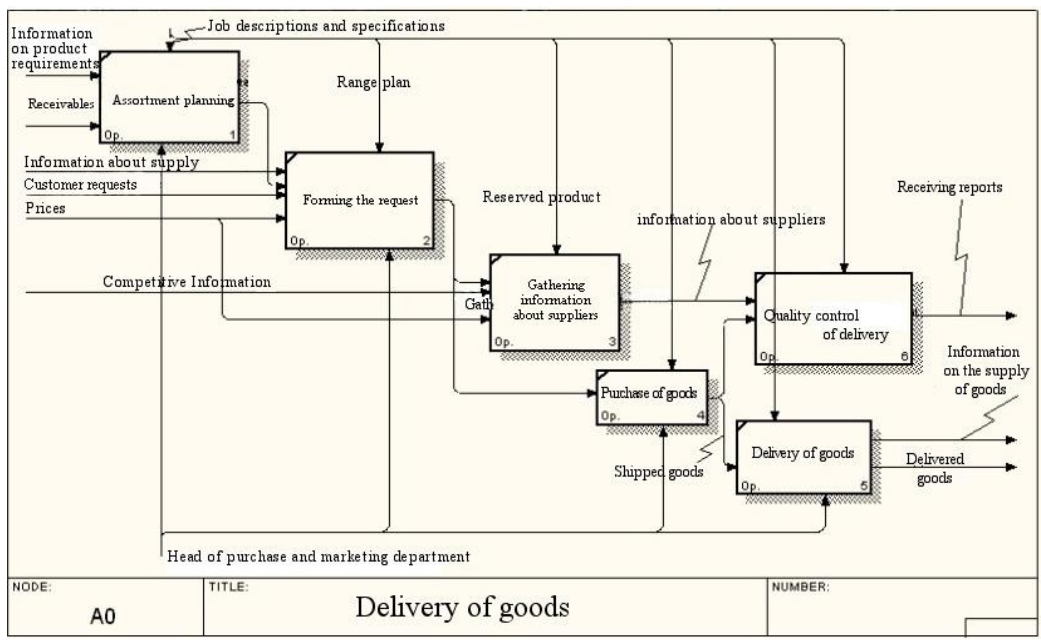

Figure 8 - business process "Delivery of goods" in violation of the sequence

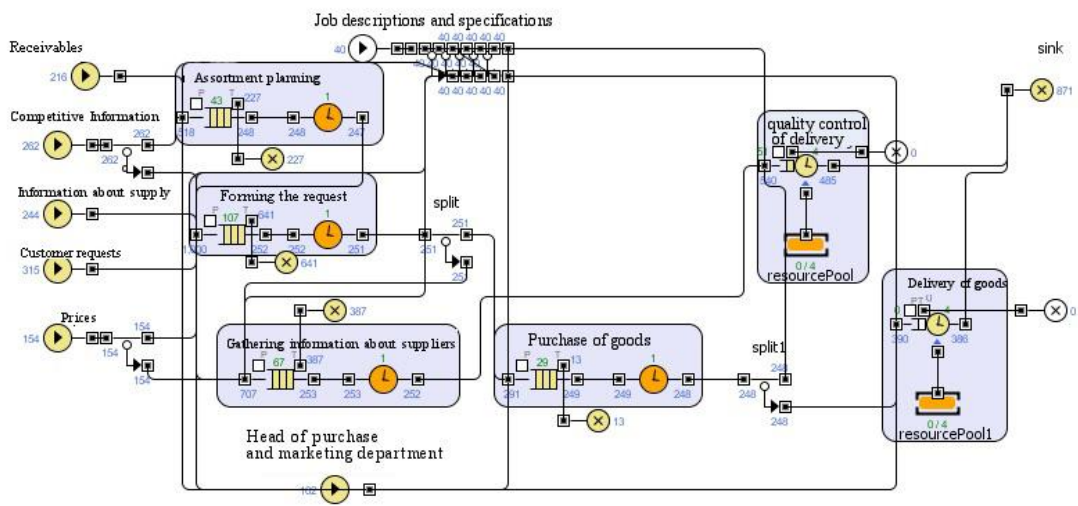

Figure 9 - the model of business process "Delivery of goods"

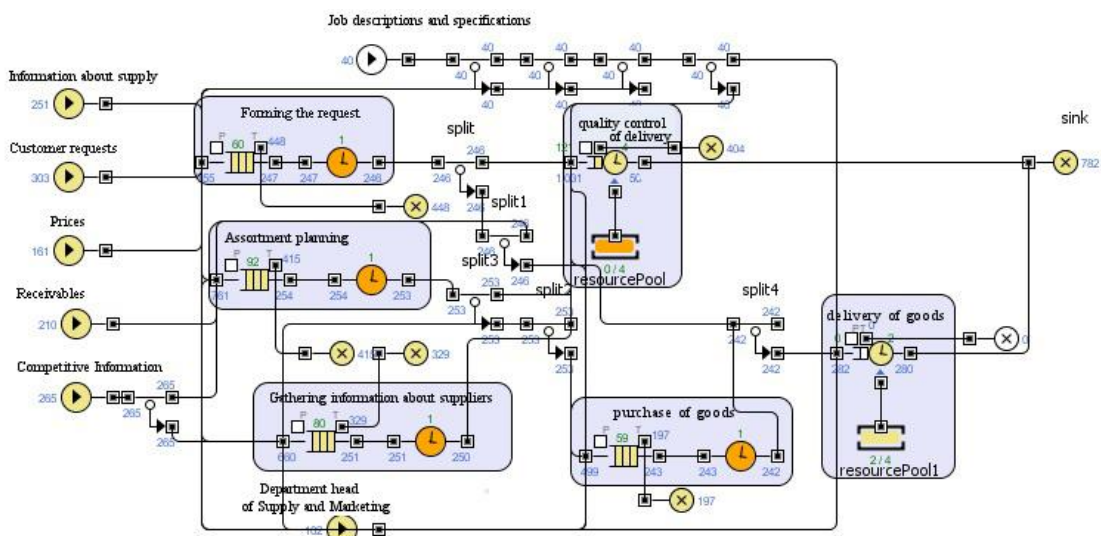

Figure 10 - modified model of business process "Delivery of goods"

Table 6 - the results of the simulation

|  | The average request time in the system | | | Utilization coefficient | | | The number of dropped requests | | |
|---|---|---|---|---|---|---|---|---|---|
|  | Until | After | % | Until | After | % | Until | After | % |
| Forming the request | 8,154 | 104,08 | Increased by 52 % | 0,997 | 0,998 | Unchanged | 414,4 | 371,4 | Decreased by 12% |
| Assortment planning |  |  |  | 0,776 | 0,998 | Increased by 28% | 103,2 | 146,6 | Increased by 42% |
| Gathering information about the suppliers |  |  |  | 0,999 | 0,996 | Unchanged | 309,6 | 306 | Decreased by 1% |
| Purchase of goods |  |  |  | 0,984 | 0,995 | Increased by 1% | 1,8 | 88,4 | Increased by 304% |
| Quality control of delivery |  |  |  | $1.818*10^{-9}$ | $1.849*10^{-9}$ | Unchanged |  | 74,6 | Increased by 375% |
| Delivery of goods |  |  |  | $1,182*10^{-9}$ | $8,816*10^{-10}$ | Decreased by $3,31*10^{-10}$ |  |  | Unchanged |

In the table 6 one can compare the simulation results of initial and modified models.

The graph below shows the change of the queue length for the simulation.

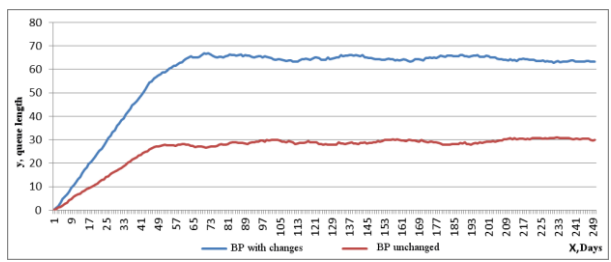

Analyzing the results of the simulation we can see that the modified process has become more complicated. Now this process requires more resources for the normal functioning.

In the modified model most requests need to be re-processed.

## V. SIMULATION (CONTINUED)

*FSBP model for parameter "Recovery time".*

In case of failure of any block or a specific functional unit, it can be some loss due to downtime.

The functional stability of the process will be better if the normal operation of the broken-down device or block will be restored faster and the probability of its failure will be lower. Also it is possible that the process will work even in case of a malfunction, but in an emergency the process will use its potential partially. In this case the restoration of normal functioning will take a little longer than it would be in a case when the operation was stopped completely. In each case, depending on the specifics of the business process, the outcome with a minimum loss is chosen.

The value of the parameter "recovery time" and several indicators (severity of failure, usage of a system during a failure) were selected in the model.

Moreover, the functional stability of the business process on the parameter "Recovery time" depends on the probability of any failure.

For simulation the business process "Product order" was chosen (Figure 11).

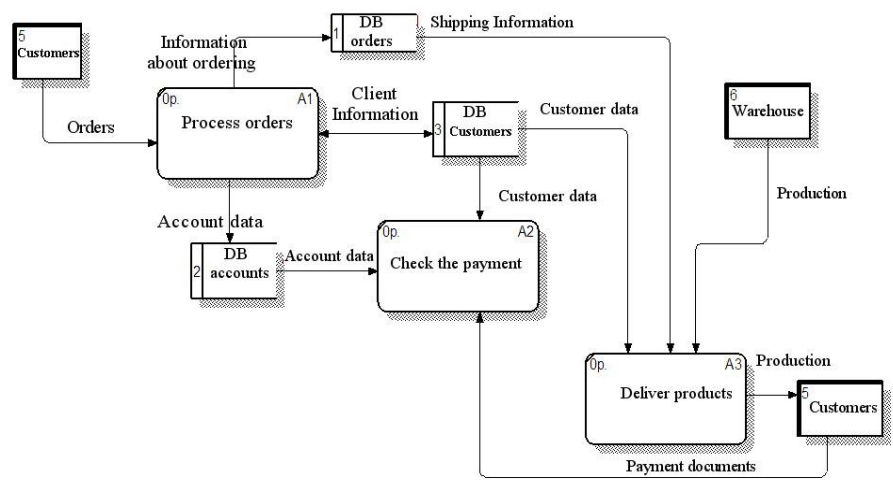

Figure 11 - The business process "Product order"

The model of the chosen business process is represented on the Figure 12.

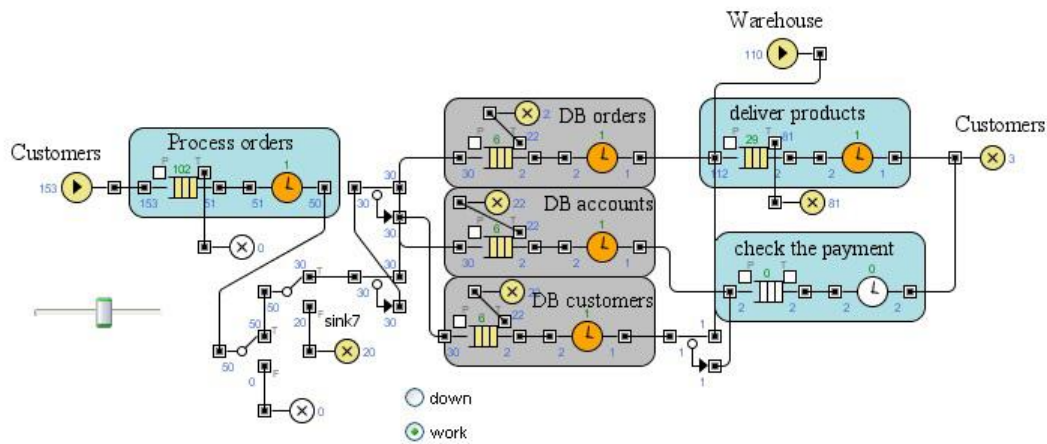

Figure 12 - The model of the business process "Product order"

For simulation the following parameters are used:

Table 7 – Sources of the model requests

| Source | μ | Description |
|---|---|---|
| Customers | 3 | Permanent, uncontrolled flow of requests that is not limited in amount. |
| Warehouse | 2 | The flow of requests is similar to the source "customers", but the amount is a little more. |

Delays are set in accordance with the chosen indicators. The graph showing the change of queue length during the simulations is represented below.

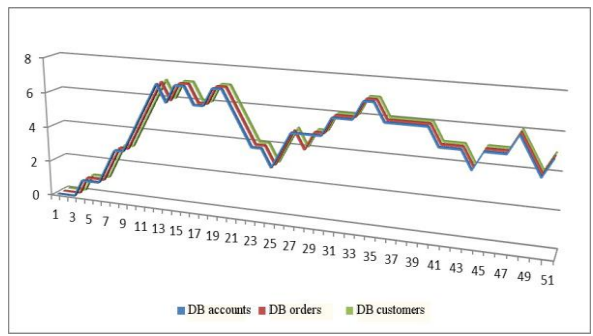

Analyzing the results of the simulation we can mention that when a disruption occurs in functional elements of the business process, the processing time of requests significantly increases, respectively increases the load of blocks. In addition the number of unprocessed, dropped by timeout requests increases also. By varying these parameters we can keep track of the most important indicators and select the optimal combination of said parameters for the particular business process.

## VI. CONCLUSION

As a result of simulation the changes of characteristics of real business process with an increased FSBP level were showcased. In this case we used the structural and the organizational parameters.

The simulation results on the legal parameters of FSBP confirm these findings.

During this work we offer, showcase and test the new approach to assessing the results of business process re-engineering by simulating their functional stability before and after re-engineering.

## REFERENCES

[1] Ray J.Paul, Vlatka Hlupic, George M. Giaglis, Simulation modelling of business processes (Brunel University, Department of Information Systems and Computing)
[2] Jon H. Weyland, Michael Engiles, Towards simulation-based business process management: *Proceedings of the 2003 Winter Simulation Conference*
[3] A. Bissay, P. Pernelle, A. Lefebvre, A. Bouras, Business processes integration and performance indicators in a PLM (LIESP – Universitй de Lyon – France)
[4] M.Abel, D.L.Shepelyansky, Google matrix of business process management (September 14,2010)